\documentclass[11pt]{article}
\usepackage{hfstyle}
\usepackage{bbm,amsmath,amssymb,wasysym,graphicx}
\usepackage{url}
\usepackage{breakurl}
\usepackage[breaklinks]{hyperref}

\usepackage[draft,firstpage]{draftmark}
\draftmarksetup{fontfamily=ptm,angle=0,scale=0.07,mark={\mbox{Published in \emph{J. of Cellular Automata}, vol. 12, no. 6, pp. 423--44 (2017)}},color=red,xcoord=-78,ycoord=120}

\newtheorem{lemma}{Lemma}[section]
\newtheorem{proposition}{Proposition}[section]

\newcommand{\ZZ}{{\mathbbm{Z}}}
\newcommand{\NN}{{\mathbbm{N}}}
\newcommand{\RR}{{\mathbbm{R}}}

\newcommand{\B}{{\mathcal B}}
\newcommand{\A}{{\mathcal A}}
\newcommand{\calS}{{\mathcal S}}
\newcommand{\f}{{\mathbf f}}

\begin{document}
\title{Explicit solution of the Cauchy problem for cellular automaton rule~172}
\author{Henryk Fuk\'s
      \oneaddress{
         Department of Mathematics\\
         Brock University\\
         \email{hfuks@brocku.ca}
       }
   }

\Abstract{
Cellular automata (CA) are fully discrete alternatives to partial differential equations (PDE).
For PDEs, one often considers the Cauchy problem, or initial value problem: find the solution
of the PDE satisfying a given initial condition. For many PDEs of the first order in time,
it is possible to find explicit formulae for the solution at the time $t>0$ if the solution is known at $t=0$.
Can something similar be achieved for CA? We demonstrate that this is indeed possible in some cases,
using elementary CA rule 172 as an example. We derive an explicit expression for the state of a given cell
after $n$ iteration of the rule 172, assuming that states of all cells are known at $n=0$. We then show
that this expression (``solution of the CA'') can be used to obtain an expected value of a given cell
after $n$ iterations, provided that the initial condition is drawn from a Bernoulli distribution.
This can be done for both finite and infinite lattices, thus providing an interesting test case for
investigating  finite size effects in CA.
}
\date{}
\maketitle

\section{Introduction}
Cellular automata are often described as fully discrete alternatives to partial differential
equations (PDEs). In one dimension, a PDE which is first-order in time can be written as
\begin{equation}\label{pde}
 u_t(x,t)=F(u,u_x,u_{xx}, \ldots),
\end{equation}
where $u(x,t)$ is the unknown function, and  the independent variables $t$ and $x$ are  commonly interpreted as, respectively, time  and position in space. Both variables $t$ and $x$, as well as $u(x,t)$, take values in the set of real numbers.

Cellular automata (CA), on the other hand, are typically written as
\begin{equation} \label{ca}
u(i,n+1)=f(u(i-r,n), u(i-r+1,n), \ldots , u(i+r,n)) ,
\end{equation}
where $f$ is called a \emph{local function} and the integer $r$ is called a \emph{radius} of the cellular 
automaton. For CA independent variables $n$ (representing time) and $i$ (representing space) are
integers, while $u(i,n)$ takes values in a finite set of symbols, usually integers.
In the case of binary cellular automata, which are the main focus of this paper,  $u(i,n)$ takes values in the set $\{0,1\}$, so that $f:\{0,1\} ^{2r+1} \rightarrow \{0,1\}$.

Comparing eqs. (\ref{pde}) and (\ref{ca}) we conclude that discrete time $n$ in CA plays the role of $t$ in PDEs, 
$i$ in CA plays the role of $x$ in PDEs, and $r$ in CA plays a similar role as the degree of the highest derivative
in PDEs. In fact, there are some further analogies, but we will not discuss them here. We will only mention that
there exist discretization schemes (such as ultradiscretization \cite{Tokihiro2004})
which allow to construct CA from PDE while preserving some features of the dynamics, but they are beyond the scope of this paper. We merely want to indicate here that conceptually, cellular automata are closely related to
PDEs, although in contrast to PDEs, all variables in CA are discrete. Moreover, dependent variable
$u$ is bounded in the case of CA -- a restriction which is not normally imposed on the dependent
variable of a PDEs.

For PDEs, the initial value problem (also called the Cauchy problem) is often
considered. It is the problem of finding $u(x,t)$ for $t>0$ subject to 
\begin{eqnarray} \label{ivppde}
 u_t(x,t)&=&F(u,u_x,u_{xx}, \ldots), \,\,\,\,\mbox{for $x \in \RR$, $t>0$},\nonumber \\
u(x,0)&=&G(x) \,\,\,\,\mbox{for $x \in \RR$},
\end{eqnarray}
where the function $G:\RR \rightarrow \RR$ represents given initial data. A similar problem can be 
formulated for cellular automata: given
\begin{eqnarray} \label{ivpca}
u(x,t+1)&=&f(u(x-r,t), u(x-r+1,t), \ldots , u(x+r,t)) \nonumber ,\\
u(x,0)&=&g(x), 
\end{eqnarray}
find $u(x,t)$ for $t>0$, where the initial data is represented by the 
function $g:\ZZ \rightarrow \{0,1\}$.

For the problem (\ref{ivpca}), it is easy to find the value of $u(x,t)$ for any
$x\in \ZZ$ and any $t\in \NN$ by direct iteration of the cellular automaton
equation (\ref{ca}). Thus, in the algorithmic sense, problem (\ref{ivpca}) is always
solvable -- all one needs to do is to take the initial data $g(x)$ and perform
$n$ iterations.

In contrast to this, the initial value problem for PDE cannot be solved exactly
by direct iteration. In some cases, however, one can obtain exact solution
in the sense of a formula for $u(x,t)$ involving $G(x)$. To give a concrete
example, consider the classical Burgers equation,
\begin{equation}
 u_t=u_{xx}+u u_x.
\end{equation}
If $u(x,0)=G(x)$, one can show that for $t>0$,
\begin{equation}
 u(x,t)=2\frac{\partial}{\partial x} \ln \left\{ 
\frac{1}{\sqrt{4 \pi t}} \int_{-\infty}^\infty \exp 
\left[ - \frac{(x-\xi)^2}{4t} -\frac{1}{2}\int_0^\xi G(\xi^\prime)d \xi^\prime \right] d \xi
\right\}.
\end{equation}
Can we obtain  similar formulae for cellular automata? The answer is affirmative \emph{in some cases}.
These cases usually involve ``simple'' CA rules. The goal of this paper is to demonstrate how 
to obtain solution of a CA in one of such ``simple'' cases, using elementary CA rule 172 as 
an example. We will also show some applications of the solution. Some ideas presented here appeared
in a preliminary form in an earlier conference proceedings paper \cite{paper39}.
\section{Basic definitions}
Let $\A=\{0,1\}$ be called {\em a symbol set}, and let $\calS =\{0,1\}^{\ZZ}$
 be the set of all bisequences over $\A$,  to be
called  {\em a configuration space}. 

{\em A block} or \emph{word} of length $n$ is an ordered set $b_{0} b_{1}
\ldots b_{n-1}$, where $n\in \NN$, $b_i \in \A$.
Let $n\in \NN$ and let
$\B_n$ denote the set of all blocks of length $n$ over $\A$ and $\B$ be
the set of all finite blocks over $\A$.

For $r \in \NN$, a mapping $f:\{0,1\}^{2r+1}\mapsto\{0,1\}$ will be called {\em a cellular
 automaton rule of radius~$r$}. Alternatively, the function $f$ can be
 considered as a mapping of $\B_{2r+1}$ into $\B_0=\A=\{0,1\}$. 

Corresponding to $f$ (also called {\em a local mapping}) we define a
 {\em global mapping}  $F:\calS \to \calS$ such that
$
(F(s))_i=f(s_{i-r},\ldots,s_i,\ldots,s_{i+r})
$
 for any $s\in \calS$.

A {\em block evolution operator} corresponding to $f$ is a mapping
 $\f:\B \mapsto \B$ defined as follows. 
Let $r\in \NN$ be the radius of $f$, and let  $a=a_0a_1 \ldots a_{n-1}\in \B_{n}$
where $n \geq 2r+1 >0$. Then 
\begin{equation}
\f(a) = \{ f(a_i,a_{i+1},\ldots,a_{i+2r})\}_{i=0}^{n-2r-1}.
\end{equation}
 Note that if
$b \in B_{2r+1}$ then $f(b)=\f(b)$.
The set of $n$-step preimages of the block $b$ under the rule $f$
is defined as the set $\f^{-n}(b)=\{ c\in \B: \f^n(c)=b\}$. Note that
the notion of block preimages has been, in somewhat different context, studied
in many earlier works, including \cite{Jen88, Jen89,Voorhees96,McIntosh2009}.

Binary rules of radius 1 are called
elementary rules, and they are usually identified
by their Wolfram number \cite{Wolfram94} $W(f)$, defined as 
\begin{equation} \label{code3}
W(f)=\sum_{x_1,x_2,x_3=0}^{1}f(x_1,x_2,x_3)2^{(2^2x_1+2^1x_2+2^0x_3)}.
\end{equation}

We will consider, as an example  of a ``solvable''
CA rule, one of the elementary rules, namely the rule with Wolfram number 172. Its local function $f: \{0,1\}^3 \to \{0,1\}$ is defined as
\begin{equation}\label{selector}
 f(x_1,x_2,x_3) = \left \{ \begin{array}{ll}
                              x_2     & \mbox{if $x_1=0$,}\\
                              x_3     & \mbox{if $x_1=1$.}
                    \end{array}
 \right.
\end{equation}
It is easy to verify that for the above $f$ we have $W(f)=172$.

The reason for which rule 172 was selected is that its dynamics is simple enough to render it
``solvable'', yet it is not entirely trivial. Further explanation regarding the meaning of ``non-trivial'' will
be given in the conclusion section. We shall also add that many properties of rule have been studied in the past,
usually in the context of other elementary CA. Some of these include
 place of rule 172 in various CA classifications, structure of its Garden of Eden configurations, algebraic properties,
and various properties of its global function 
\cite{Martinez2013,Bulitko2006,Voorhees93,Nishio2010}.

In what follows, whenever we use the symbol $f$ it will signify the local function defined in eq. 
(\ref{selector}), while $\f$ and $F$ will denote, respectively,
the corresponding block evolution operator and the global function. To familiarize the reader with the concept of the block evolution
operator, let us take as an example $b=1001010$. We can compute $\f(b)$ by applying $f$ to all consecutive
triples of symbols, that is,  $\f(b)=f(100) f(001) f(010) f(101) f(010)=00111$. If we  apply $\f$ again to $00111$,
we will obtain $\f^2(b)=011$, and yet another application of $\f$ yields  $\f^3(b)=1$. It is sometimes convenient to write
consecutive images of $b$ under each other, as follows:
 \begin{verbatim}
           1001010
            00111
             011
              1
\end{verbatim}
The above shows, starting from the top, $b$, $\f(b)$, $\f^2(b)$, and $\f^3(b)$. 

We shall also note that there is usually more than one block $c$ such that $\f(b)=c$. For example, for rule 172, 
 $\f(0010)=\f(0011)=\f(1101)=01$. We can, therefore, write $\f^{-1}(01)=\{0010,0011,1101\}$.  Similarly, we can write
\begin{equation}
\f^{-2}(101)=\{
0011101, 
0101101, 
0111101, 
1011101, 
1101101, 
1111101\},
\end{equation}
because all 6 block on the right hand side of the above (and only these blocks) have the property that after applying $\f$ to them twice,
one obtains $101$.

Our strategy for constructing the solution of rule 172 will be as follows. First, we will construct $n$-step preimages
of $1$ (i.e., sets $\f^{-n}(1)$) for various $n$. 
We will then try to find patterns in these set which would allow us
to give a combinatorial description of them, as set of binary strings satisfying certain conditions. Once
this is done, we will construct a Boolean function which is an indicator function of 
$\f^{-n}(1)$.  Such function will then be used to construct an explicit expression for $[F^n(x)]_i$ for any $x\in \{0,1\}^{\ZZ}$,
$i \in \ZZ$ and $n \in \NN$.
\section{Structure of preimage sets}
Suppose now that we have a string $b$ of length $2n+1$ and we want to find out the necessary and sufficient
conditions for $\f^n(b)=1$. We will try to ``guess'' these conditions first, formulate them in a rigorous way, and then
prove them.

\begin{figure}
  \begin{center}
    \includegraphics[scale=0.5]{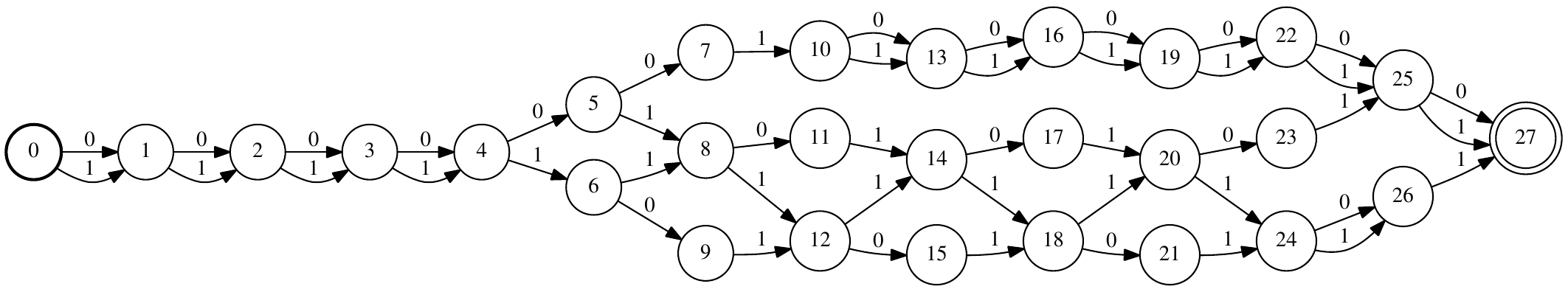}
  \end{center}
 \caption{Finite state machine producing $\f^{-6}(1)$ for elementary CA rule 172.}
  \label{fsmexample}
\end{figure}

In order to ``guess'' the conditions, one can generate sets $\f^{-n}(1)$ for various values of $n$ and try to 
discover obvious patters in them. From author's experience, a good way to do this is to build minimal finite state machines
generating words of $\f^{-n}(1)$.  This can be done using AT\&T FSM Library~\cite{fsmlib,Skiena2008}, and Figure~\ref{fsmexample} shows
an example of a minimal finite state machine (FSM) generating $\f^{-7}(1)$ for rule 172.
In order to generate a  preimage of 1 using this picture, start on the left (at circled zero) and 
follow the arrows writing down all encountered edge labels  until you reach the final state (doubly circled 27).
The string of 15 labels obtained this way will be a possible preimage of $1$, one of many. Obviously there are as many preimage
string as paths joining the initial state and the final state. Note that circled numbers denote 
internal states of the FSM, and are irrelevant for our purposes.

From Figure~\ref{fsmexample}, it is clear that the first 5 symbols of $\f^{-6}(1)$ are arbitrary,
and then we have two possibilities:
\begin{itemize}
 \item[(i)] 001 followed by 6 arbitrary symbols, or
 \item[(i)] string of 8 symbols without 00 pair anywhere, followed by 10 or 11 (if it ends with 0)
or by 01 or 11 (if it ends by 1).
\end{itemize}
This observation can be generalized and  summarized as the following proposition.
\begin{proposition} \label{main}
  Block $b$ of length $2n+1$ belongs to $\f^{-n}(1)$ if and only if it has the structure 
 \begin{equation}
  b= \underbrace{\star \star \ldots \star}_{n-2} 001 \underbrace{\star \star \ldots \star}_{n},
\end{equation}
or
 \begin{equation}
  b= \underbrace{\star \star \ldots \star}_{n-2}  a_1 a_2\ldots a_{n+1}c_1c_2,
\end{equation}
where $ a_1 a_2\ldots a_n$ is a binary string which does not contain any pair of adjacent zeros,
 and
\begin{equation} \label{condc1c2}
 c_1c_2= 
\begin{cases}
1 \star , & \text{if}\,\,\, a_{n+1}=0, \\
\star 1, & \text{otherwise}.
\end{cases}
\end{equation}
\end{proposition}
We will sketch the proof of the above, leaving out some tedious details.
 \begin{figure}
  \begin{center}
    \includegraphics[scale=0.6]{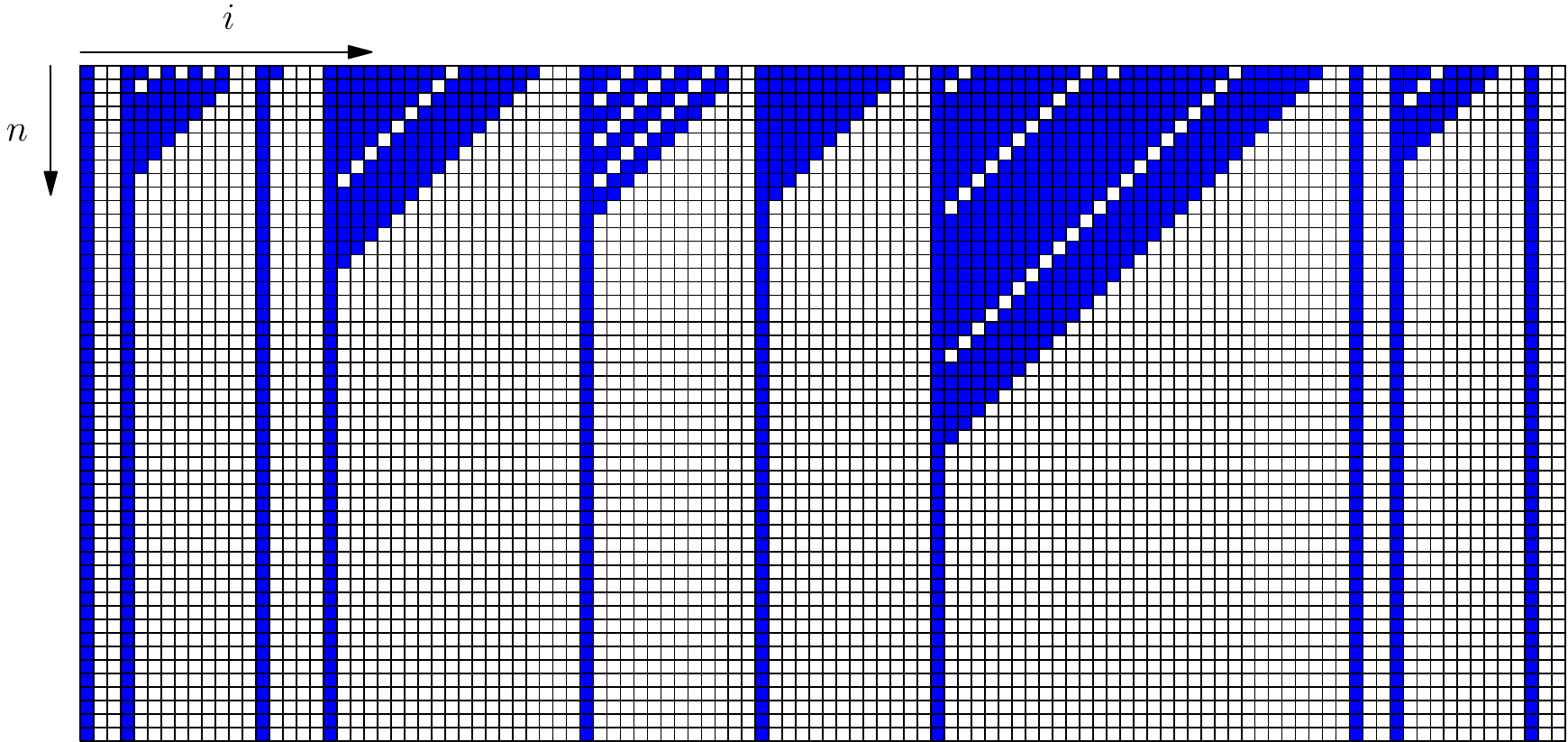}
  \end{center}
 \caption{Example of a spatiotemporal pattern produced by rule 172.}
  \label{figpattern}
\end{figure}
 It will be helpful to inspect spatiotemporal pattern generated by
rule 172 first, as shown in Figure~\ref{figpattern}. Careful inspection of this pattern
reveals three facts, each of them easily provable in a rigorous way:
\begin{itemize}
 \item[(F1)] A cluster of two or more zeros keeps its right boundary 
in the same place for ever.
 \item[(F2)] A cluster of two or more zeros extends its left boundary to the left one unit 
per time step as long as the left boundary is preceded by two
or more ones. If the left boundary of the cluster of zeros is  01, the cluster does not grow.
 \item[(F3)] Isolated zero moves to the left one step at a time as long as it has at least two ones on the left.
If an isolated zero is preceded by 01, it disappears in the next time step.
\end{itemize}

Suppose now that we have a string $b$ of length $2n+1$ and we want to find out the necessary and sufficient
conditions for $\f^n(b)=1$. From (F1) it is clear that the word $001$ will remain in the same position forever, 
which means that if  
\begin{equation}
  b= \underbrace{\star \star \ldots \star}_{n-2} 001 \underbrace{\star \star \ldots \star}_{n},
\end{equation}
then  $\f^n(b)=1$. What are the other possibilities for $b$ which would result in $\f^n(b)=1$? 

From (F2) we deduce that if there is no cluster of two or more more zeros somewhere
in the last $n+3$ bits of $b$, then there is no possibility of the growth of cluster of zeros
producing $\f^n(b)=0$. The only way to get zero after $n$ iterations of $\f$ in such a case
would be having zero at the end of $b$ preceded by $11$. This means that in order to avoid this scenario,
the last 3 bits of $b$ must be $010$, $011$, $101$ or $111$, or, in other words, the last three bits
must be $01\star$ or $1\star1$, as in eq.~(\ref{condc1c2}). $\Box$

\section{Solving rule 172}
 We are now almost ready to construct the solution of rule 172. Suppose that $x \in \{0, 1\}^{\ZZ}$ is an initial configuration,
 and that we iterate rule 172 $n$ times. What is the value of the central site after $n$ iterations of the rule, that is, the value of $[F^n(x)]_0$? Obviously it can be either 0 or 1, so let us suppose that
it is 1, which means that $x_{-n} x_{-n+1}, \ldots x_n \in \f^{-n}(1)$. By the virtue of Proposition \ref{main},
$x_{-n} x_{-n+1}, \ldots x_n $ 
must take one of the two forms, the fist of them being
\begin{equation}
  x_{-n} x_{-n+1}, \ldots x_n =\underbrace{\star \star \ldots \star}_{n-2} 001 \underbrace{\star \star \ldots \star}_{n}.
\end{equation}
The above means that $x_{-2}x_{-1}x_0=001$, and it will be true if and only if 
\begin{equation} \label{pos1}
 (1-{x}_{-2})(1-{x}_{-1})x_0=1.
\end{equation}
The second possibility, according to Proposition \ref{main}, is
\begin{equation}
  x_{-n} x_{-n+1}, \ldots x_n =\underbrace{\star \star \ldots \star}_{n-2}  a_1 a_2\ldots a_{n+1}c_1c_2,
\end{equation}
where $a_ia_{i+1}\neq 00$ for $i = 1,2,\ldots, n$ and $c_1$, $c_2$ satisfy condition of eq. (\ref{condc1c2}). This means that
$a_1a_2 \ldots a_{n+1}=c_{-2}c_{-1}\ldots c_{t-2}$, and therefore $x_{i}x_{i+1}\neq 00$ for $i=-2, -1, \ldots, t-3$, as well as
\begin{equation} 
 x_{t-1}x_t=  
\begin{cases} 
1 \star , & \text{if}\,\,\, x_{t-2}=0, \\
\star 1, & \text{otherwise}.
\end{cases}
\end{equation}
 The second possibility will be realized if and only if
\begin{equation} \label{pos2}
 \left( \prod_{i=-2}^{t-3}(1-\bar{x}_i\bar{x}_{i+1}) \right)
(\bar{x}_{t-2}x_{t-1} + x_{t-2}x_t)=1,
\end{equation}
where we used notation $\bar{x}_i=1-x_1$. Combining eqs. (\ref{pos1}) and (\ref{pos2}) we obtain
\begin{equation} \label{sol0}
 [F^n(x)]_0=\bar{x}_{-2}\bar{x}_{-1}x_0 +  \left( \prod_{i=-2}^{n-3}(1-\bar{x}_i\bar{x}_{i+1}) \right)
(\bar{x}_{n-2}x_{n-1} + x_{n-2}x_n).
\end{equation}
This is the desired solution expressing the value of the central site after $n$ iterations of rule $F$ starting from an initial 
configuration $x$. Of course, we can now obtain analogous expression for any other site $[F^n(x)]_j$, by simply translating the above
formula from $j=0$ to an arbitrary position $j$. 
\begin{proposition}
 Let $F$ be the global function of elementary CA rule 172 and $x \in \{0,1\}^{\ZZ}$. Then, after $n \in \NN$ iterations
of $F$, for any $j\in \ZZ$,
\begin{multline} \label{solj}
[F^n(x)]_j=\bar{x}_{j-2}\bar{x}_{j-1}x_j \\+  \left( \prod_{i=j-2}^{j+n-3}(1-\bar{x}_{i}\bar{x}_{i+1}) \right)
(\bar{x}_{j+n-2}x_{j+n-1} + x_{j+n-2}x_{j+n}).
\end{multline}
\end{proposition}
The above could be called a solution of rule 172. It is an explicit solution of the
Cauchy problem for this rule, expressing the state of a site at position $j$ after $n$ iterations
in terms of initial site values. The formula for the solution is very simple, utilizing only addition,
subtraction, and multiplication of site values. As we will see in subsequent sections, it can be
very useful in practice, for example for constructing probabilistic solutions of the CA rule and investigating
finite size effects.
\section{Probabilistic solution for infinite configurations}
Let us now assume that the initial configuration $x$ is drawn from a Bernoulli distribution. More precisely, let
$x_j=X_j$ for $j \in \ZZ$, where  $X_j$ are independent and identically distributed random variables 
such that  $Pr(X_j=1)=q$, $Pr(X_j=0)=1-q$, where $q \in [0,1]$. What is the expected value of 
$[F^n(x)]_j$ in such circumstances? Denoting the expected value by $\langle \cdot \rangle$, let us first note that
due to translational invariance, $\langle [F^n(x)]_j \rangle=\langle [F^n(x)]_0 \rangle$. In order to compute
$\langle [F^n(x)]_0 \rangle$, we need to calculate the expected value of the right hand side of eq. (\ref{sol0}).
The following lemma will be useful for this purpose.
\begin{lemma} \label{prodlemma}
 Let $q \in (0,1)$ and let $X_i$ be independent and identically distributed Bernoulli random variables for $i \in \{0, 1, \ldots, n\}$
such that  $Pr(X_i=1)=q$, $Pr(X_i=0)=1-q$. Then
\begin{equation} \label{expectedproduct}
 \left \langle \prod_{i=1}^{n-1} \left(1-\bar{X}_i \bar{X}_{i+1} \right) \right \rangle=
\frac{q}{\lambda_2-\lambda_1} \left( \alpha_1 \lambda_1^{n-2} + \alpha_2 \lambda_2^{n-2} \right),
\end{equation}
where
\begin{align}
 \lambda_{1,2}&=\frac{1}{2}q \pm \frac{1}{2}\sqrt{q(4-3q)},\label{eigenvalues}\\
\alpha_{1,2}&=   \left( \frac{q}{2}-1 \right) \sqrt {q \left( 4-3\,q \right) } \pm  \left( \frac{q^2}{2} -1 \right).
\end{align}
\end{lemma}

To prove it, let us first define
\begin{equation}
U_n=\prod_{i=1}^{n-1} \left(1-\bar{X}_i \bar{X}_{i+1} \right) \mbox{\,\,\,\, and \,\,\,\,}
V_n=\bar{X}_n\prod_{i=1}^{n-1} \left(1-\bar{X}_i \bar{X}_{i+1} \right). 
\end{equation}
We observe that
\begin{equation}
 U_n=U_{n-1}(1-\bar{X}_{n-1} \bar{X}_n)=U_{n-1} - \bar{X}_n V_{n-1},
\end{equation}
and
\begin{equation}
 V_n=\bar{X}_n U_n=\bar{X}_n U_{n-1}(1-\bar{X}_{n-1} \bar{X}_n)=\bar{X}_n U_{n-1} -\bar{X}_n V_{n-1},
\end{equation}
where we used the fact $X_n$ is a Boolean variable, thus  $\bar{X}_n^2=\bar{X}_n$. This yields
the system of recurrence equations for $U_n$ and $V_n$,
\begin{align} \label{recuV}
  U_n&=U_{n-1} - \bar{X}_n V_{n-1},\\
   V_n&=\bar{X}_n U_{n-1} -\bar{X}_n V_{n-1}. \nonumber
\end{align}
Since $\bar{X}_n$ is independent of both $U_{n-1}$ and $V_{n-1}$, we can write
\begin{align}
  \langle U_n \rangle&=\langle U_{n-1}\rangle - \langle\bar{X}_n\rangle \langle V_{n-1}\rangle,\\
   \langle V_n \rangle&=\langle \bar{X}_n\rangle \langle U_{n-1}\rangle -\langle \bar{X}_n\rangle \langle V_{n-1}\rangle.
\end{align}
Now, taking into account that $\langle \bar{X}_n\rangle = 1-q$, we obtain
\begin{equation}
 \left[ \begin {array}{c} 
\langle U_n \rangle\\
\langle V_n \rangle
 \end {array} \right]
=M
\left[ \begin {array}{c} 
\langle U_{n-1} \rangle\\
\langle V_{n-1} \rangle
 \end {array} \right],
\end{equation}
where
\begin{equation}
 M=
\left[ \begin {array}{cc} 
1 & q-1\\
1-q & q-1
 \end {array} \right].
\end{equation}
This recurrence equation is easy to solve,
\begin{equation}
 \left[ \begin {array}{c} 
\langle U_n \rangle\\
\langle V_n \rangle
 \end {array} \right]
=M^{n-2}
\left[ \begin {array}{c} 
\langle U_{2} \rangle\\
\langle V_{2} \rangle
 \end {array} \right].
\end{equation}
Since  $\langle U_{2} \rangle$ and $\langle V_{2} \rangle$ can be directly computed,
\begin{equation}
 \langle U_{2} \rangle=\langle 1-\bar{X}_1 \bar{X}_2 \rangle=1-(1-q)^2=2q-q^2,
\end{equation}
\begin{equation}
 \langle V_{2} \rangle=\langle \bar{X}_2( 1-\bar{X}_1 \bar{X}_2 )\rangle=\langle \bar{X}_2 -\bar{X}_1 \bar{X}_2 \rangle
=1-q - (1-q)^2=q-q^2,
\end{equation}
we obtain
\begin{equation}
 \left[ \begin {array}{c} 
\langle U_n \rangle\\
\langle V_n \rangle
 \end {array} \right]
=
M^{n-2}
\left[ \begin {array}{c} 
2q-q^2\\
q-q^2 
 \end {array} \right].
\end{equation}
The only thing left is to compute is  $M^{n-2}$. This can be done by diagonalizing $M$,
\begin{equation}
M^{n-2}=
P
\left[ \begin {array}{cc} \lambda_1^{n-2}&0
\\ \noalign{\medskip}0&\lambda_2^{n-2}\end {array} \right] P^{-1},
\end{equation}
where  $\lambda_{1,2}$ are eigenvalues of $M$, as defined in eq. (\ref{eigenvalues}), and $P$
is the matrix of eigenvectors of $P$, 
\begin{equation} \label{defP}
P=\left[ \begin {array}{cc} {\frac {1-q}{1-\lambda_{{1}}}}&{\frac {1-
q}{1-\lambda_{{2}}}}\\ \noalign{\medskip}1&1\end {array} \right], 
\mbox{\,\,\,}
 P^{-1}=
\frac{1}{\lambda_1-\lambda_2}
 \left[ \begin {array}{cc} -{\frac { \left( \lambda_{{1}}-1 \right) 
 \left( \lambda_{{2}}-1 \right) }{-1+q}}&\lambda_{{1}}-1
\\ \noalign{\medskip}{\frac { \left( \lambda_{{1}}-1 \right)  \left( 
\lambda_{{2}}-1 \right) }{-1+q}}&-\lambda_{{2}}+1\end {array} \right].
\end{equation} \label{fialUV}
The final formula for $U_n$ and $V_n$ is
\begin{equation} \label{finalUV}
 \left[ \begin {array}{c} 
\langle U_n \rangle\\
\langle V_n \rangle
 \end {array} \right]
=P
\left[ \begin {array}{cc} \lambda_1^{n-2}&0
\\ \noalign{\medskip}0&\lambda_2^{n-2}\end {array} \right] P^{-1}
\left[ \begin {array}{c} 
2q-q^2\\
q-q^2 
 \end {array} \right].
\end{equation}
By carrying out the multiplications of matrices on the right hand side, after some algebra, we obtain
\begin{multline}
\langle U_n \rangle= 
\frac{q}{\lambda_2-\lambda_1}
 \Big(  \left( -1-q+{q}^{2}+2\,\lambda_{{2}}-q\lambda_{{2}}
 \right) \lambda_1^{n-2}\\
+ \left( 1+q-{q}^{2}-2\,\lambda_{{1}}+q\lambda_{
{1}} \right) \lambda_2^{n-2} \Big) ,
\end{multline}
in agreement with eq. (\ref{expectedproduct}). \Square

Before we take the expected value of both sides of eq. (\ref{sol0}), let us first rewrite the last factor,
\begin{multline}
 \bar{x}_{n-2 }x_{n-1} + x_{n-2} x_n \\
=(1-x_{n-2}) x_{n-1} + x_{n-2}x_n=x_{n-1}+ x_{n-2}(x_n-x_{n-1}).
\end{multline}
Using the above, we obtain from eq. (\ref{sol0}),
\begin{multline} 
\langle [F^n(x)]_0 \rangle= \langle \bar{x}_{-2}\bar{x}_{-1}x_0 \rangle 
+ \left\langle \left( \prod_{i=-2}^{n-3}(1-\bar{x}_i\bar{x}_{i+1}) \right)
x_{n-1} \right\rangle \\
+ \left\langle   \left( \prod_{i=-2}^{n-3}(1-\bar{x}_i\bar{x}_{i+1}) \right)
x_{n-2}(x_n-x_{n-1}) \right\rangle .
\end{multline}
Using the fact that the expected value of the product of independent random variables is equal to the product of their
expected values, this yields
\begin{multline} 
\langle [F^n(x)]_0 \rangle= \langle \bar{x}_{-2}\rangle  \langle \bar{x}_{-1}\rangle  \langle x_0 \rangle 
+ \left\langle  \prod_{i=-2}^{n-3}(1-\bar{x}_i\bar{x}_{i+1}) 
 \right\rangle \langle x_{n-1} \rangle \\
+ \left\langle   \left( \prod_{i=-2}^{n-3}(1-\bar{x}_i\bar{x}_{i+1}) \right) x_{n-2} \right\rangle 
 \langle x_n-x_{n-1} \rangle.
\end{multline}
Because $\langle x_n-x_{n-1} \rangle=0$, the last term vanishes, and, using the fact that $\langle{x}_{i}\rangle=q$ and 
$\langle \bar{x}_{i}\rangle=1-q$, we obtain
\begin{equation} 
\langle [F^n(x)]_0 \rangle= (1-q)^2 q
+ q \left\langle  \prod_{i=-2}^{n-3}(1-\bar{x}_i\bar{x}_{i+1}) 
 \right\rangle .
\end{equation}
Using Lemma \ref{prodlemma}, and remembering that the expected value must be the same for any index $j$, the final result is thus
\begin{equation} \label{infinitesolution}
\langle [F^n(x)]_j \rangle= (1-q)^2 q
+\frac{q^2}{\lambda_2-\lambda_1} \left( \alpha_1 \lambda_1^{n-1} + \alpha_2 \lambda_2^{n-1} \right).
\end{equation}
The above could be called \emph{a probabilistic solution} of CA rule 172 for infinite Bernoulli initial configuration.
Note that since $|\lambda_{1,2}|<1$, we have
\begin{equation}
 \lim_{n \to \infty} \langle [F^n(x)]_j \rangle =(1-q)^2 q.
\end{equation}

When $q=1/2$, eq. (\ref{infinitesolution}) becomes, after simplification,
\begin{equation} 
\langle [F^n(x)]_j \rangle=\frac{1}{8}+ 
\left(  \frac{1}{4}+\frac{\sqrt {5}}{10} \right)  \left(  \frac{1}{4} + \frac{\sqrt {5}}{4}  \right) ^{n}+ 
\left( \frac{1}{4}-\frac{\sqrt {5}}{10} \right)  \left(  \frac{1}{4}- \frac{\sqrt {5}}{4} \right) ^{n}.
\end{equation}
Since $\frac{1}{4} + \frac{\sqrt {5}}{4}$ is half of \emph{ratio divina} (the golden ratio), one recognizes
a link to Fibonacci numbers in the above. Indeed, it is easy to show that for $q=1/2$,
\begin{equation}
\langle [F^n(x)]_j \rangle =\frac{1}{8}  + \frac{{\cal F}_{n+3}}{2^{n+2}},
\end{equation}
where ${\cal F}_{n}$ is the $n$-th Fibonacci number. 
\section{Probabilistic solution for periodic configuration}
Suppose now that the initial condition is periodic with period $k$, that is, $x_i=x_{i+k}$ for all $i \in \ZZ$. 
Although one could of course consider all finite configurations of a given length and determine their attractors,
it will nevertheless be useful to construct a general formula valid for arbitrary $k$. This could be, for example, 
useful if one wants  to study the dependence of the speed of convergence to the steady state as a function of $k$.

We will take $i=0, \ldots k-1$ as the principal period. 
The solution will be given by the same formula as before, except that all indices are to be taken modulo $k$. 
Let us further assume that, as before,  $x_j=X_j$ for $j \in \{0,1, \ldots, k-1\}$, where  $X_j$ are independent and identically distributed random variables  such that  $Pr(X_j=1)=q$, $Pr(X_j=0)=1-q$, where $q \in [0,1]$.

In order to compute the expected value of a site after $n$ iterations of rule 172, we take expected value of both sides of eq. (\ref{solj})
(remembering that indices are now modulo $k$), obtaining
\begin{multline}
\langle [F^n(x)]_j \rangle  =(1-q)^2 q \\
 +  \left \langle \left( \prod_{i=j-2}^{j+n-3}(1-\bar{x}_{i}\bar{x}_{i+1}) \right)
(\bar{x}_{j+n-2}x_{j+n-1} + x_{j+n-2}x_{j+n}) \right \rangle.
\end{multline}
Observe that when  $j=-2$ (remember that the expected value must be $j$-independent, so the choice of $j$ does not matter),
the only indices of $x$ occurring on the right hand side of the above will be in the range from 0 to $n+2$. This means that for
for $n \leq k-3$, we will never actually need to use modulo $k$ operation to bring the index to the
principal period range. For this reason, eq. (\ref{infinitesolution}) remains valid in the periodic case as long as 
$n \leq k-3$.

Let us now suppose that $n \geq k$. In this case,
\begin{equation}
 \prod_{i=j-2}^{j+n-3}(1-\bar{x}_{i}\bar{x}_{i+1}) = \prod_{i=0}^{k-1}(1-\bar{x}_{i}\bar{x}_{i+1}) ,
\end{equation}
because in the product on the left hand side there are only $k$ different factors, and $(1-\bar{x}_i \bar{x}_{i+1})^m =(1-\bar{x}_i \bar{x}_{i+1})$
for any positive integer $m$. We will once again take advantage of the translational symmetry. Since 
$\langle [F^n(x)]_j \rangle$ should be the same for all $j$,  we will take $j=k-n$, 
\begin{multline}
\langle [F^n(x)]_j \rangle =\langle [F^n(x)]_{k-n} \rangle=(1-q)^2 q + \\ \left \langle \left( \prod_{i=0}^{k-1}(1-\bar{x}_{i}\bar{x}_{i+1}) \right)
(\bar{x}_{k-2}x_{k-1} + x_{k-2}x_{0}) \right \rangle.
\end{multline}
Before we proceed further, let us note that
\begin{multline}
\bar{x}_{k-2}x_{k-1} + x_{k-2}x_{0}=
 \bar{x}_{k-2}(1-\bar{x}_{k-1}) + (1-\bar{x}_{k-2})(1-\bar{x}_{0}) \\=
1-\bar{x}_{k-2}\bar{x}_{k-1} -\bar{x}_0 + \bar{x}_0 \bar{x}_{k-2}.
\end{multline}
Since $(1-\bar{x}_{k-2}\bar{x}_{k-1}) \prod_{i=0}^{k-1}(1-\bar{x}_{i}\bar{x}_{i+1})=\prod_{i=0}^{k-1}(1-\bar{x}_{i}\bar{x}_{i+1})$,
we obtain
\begin{multline} \label{twoexp}
\langle [F^n(x)]_j \rangle =(1-q)^2 q +  
\left \langle  \prod_{i=0}^{k-1}(1-\bar{x}_{i}\bar{x}_{i+1})  \right \rangle \\
+\left \langle \bar{x}_0 (\bar{x}_{k-2}-1) \prod_{i=0}^{k-1}(1-\bar{x}_{i}\bar{x}_{i+1})  \right \rangle.
\end{multline}
We will deal with the two expected values on the right hand side separately. Let us start from the first one.
\begin{gather*}
 \left \langle \prod_{i=0}^{k-1} (1-\bar{x}_i\bar{x}_{i+1})\right \rangle
= \left \langle (1-\bar{x}_0 \bar{x}_1) \prod_{i=1}^{k-2} (1-\bar{x}_i\bar{x}_{i+1}) (1-\bar{x}_{k-1} \bar{x}_0) \right \rangle\\
= \left \langle (1-\bar{x}_0 \bar{x}_1) (1-\bar{x}_{k-1} \bar{x}_0) \prod_{i=1}^{k-2} (1-\bar{x}_i\bar{x}_{i+1})  \right \rangle\\
= \left \langle (1-\bar{x}_0 \bar{x}_{k-1}-\bar{x}_0 \bar{x}_{1}+\bar{x}_0 \bar{x}_{1} \bar{x}_{k-1}) \prod_{i=1}^{k-2} (1-\bar{x}_i\bar{x}_{i+1})  \right \rangle  \\
=\left \langle \prod_{i=1}^{k-2} (1-\bar{x}_i\bar{x}_{i+1})  \right \rangle  
- \langle \bar{x}_0 \rangle \left \langle  \bar{x}_{k-1}  \prod_{i=1}^{k-2} (1-\bar{x}_i\bar{x}_{i+1})  \right \rangle \\
- \langle \bar{x}_0 \rangle  \left \langle  \bar{x}_1  \prod_{i=1}^{k-2} (1-\bar{x}_i\bar{x}_{i+1})  \right \rangle 
+ \langle \bar{x}_0 \rangle  \left \langle   \bar{x}_1 \bar{x}_{k-1} \prod_{i=1}^{k-2} (1-\bar{x}_i\bar{x}_{i+1})  \right \rangle.  
\end{gather*}
Recall that 
\begin{equation}
U_n=\prod_{i=1}^{n-1} \left(1-\bar{X}_i \bar{X}_{i+1} \right) \mbox{\,\,\,\, and \,\,\,\,}
V_n=\bar{X}_n\prod_{i=1}^{n-1} \left(1-\bar{X}_i \bar{X}_{i+1} \right),
\end{equation}
and define
\begin{equation}
U^{\prime}_n=\bar{X}_1\prod_{i=1}^{n-1} \left(1-\bar{X}_i \bar{X}_{i+1} \right) \mbox{\,\,\,\, and \,\,\,\,}
V^{\prime}_n=\bar{X}_1 \bar{X}_n\prod_{i=1}^{n-1} \left(1-\bar{X}_i \bar{X}_{i+1} \right).
\end{equation}
Now we have
\begin{multline}
 \left \langle \prod_{i=0}^{k-1} (1-\bar{x}_i\bar{x}_{i+1})\right \rangle
=\langle U_{k-1} \rangle 
-(1-q) \langle V_{k-1} \rangle
-(1-q) \langle U^{\prime}_{k-1} \rangle \\
+(1-q) \langle V^{\prime}_{k-1} \rangle
=\langle U_{k} \rangle + q \langle U^{\prime}_{k-1} \rangle 
- \langle U^{\prime}_{k-1} \rangle
+(1-q) \langle V^{\prime}_{k-1} \rangle \\
=\langle U_{k} \rangle + q \langle U^{\prime}_{k-1} \rangle  - \langle U^{\prime}_{k} \rangle ,
\end{multline}
where we used the fact that $\langle U^{\prime}_{n} \rangle, \langle V^{\prime}_{n} \rangle$
satisfy the same recurrence equations as  $\langle U_{n} \rangle, \langle V_{n} \rangle$.

The second expected value in eq. (\ref{twoexp}) can be calculated as follows.
\begin{multline*}
 \left \langle \bar{x}_0 (\bar{x}_{k-2}-1) \prod_{i=0}^{k-1}(1-\bar{x}_{i}\bar{x}_{i+1})  \right \rangle \\
=\left \langle \bar{x}_0 (\bar{x}_{k-2}-1)(1-\bar{x}_{k-2}\bar{x}_{k-1}) (1-\bar{x}_{k-1}\bar{x}_{0}) \prod_{i=0}^{k-3}(1-\bar{x}_{i}\bar{x}_{i+1}) 
 \right \rangle \\
= \left \langle  \bar{x}_0 (\bar{x}_{{k-2}}-\bar{x}_{{k-2}}\bar{x}_{{k-1}}-1+\bar{x}_{{k-1}})
\prod_{i=0}^{k-3}(1-\bar{x}_{i}\bar{x}_{i+1}) 
 \right \rangle=\\
 \left \langle  \bar{x}_0 \bar{x}_{{k-2}}
\prod_{i=0}^{k-3}(1-\bar{x}_{i}\bar{x}_{i+1}) \right \rangle +
 \left \langle  \bar{x}_0 (1-\bar{x}_{{k-2}}\bar{x}_{{k-1}})
\prod_{i=0}^{k-3}(1-\bar{x}_{i}\bar{x}_{i+1}) \right \rangle
\\ -  2\left \langle  \bar{x}_0 
\prod_{i=0}^{k-3}(1-\bar{x}_{i}\bar{x}_{i+1}) \right \rangle \\
+ \left \langle  \bar{x}_0 \bar{x}_{k-1}
\prod_{i=0}^{k-3}(1-\bar{x}_{i}\bar{x}_{i+1}) \right \rangle=
\langle V^{\prime}_{k-1}\rangle + \langle  U^{\prime}_k \rangle-2 \langle U^{\prime}_{k-1}\rangle+ (1-q) \langle U^{\prime}_{k-1}\rangle .
\end{multline*}
Combining both expected values computed above we get
\begin{multline}
\langle [F^n(x)]_j \rangle =(1-q)^2 q +
 \langle U_{k} \rangle + q \langle U^{\prime}_{k-1} \rangle  - \langle U^{\prime}_{k} \rangle 
+
\langle V^{\prime}_{k-1}\rangle + \langle  U^{\prime}_k \rangle \\
-2 \langle U^{\prime}_{k-1}\rangle+ (1-q) \langle U^{\prime}_{k-1}\rangle \nonumber 
=(1-q)^2 q + \langle U_{k} \rangle -\langle U^{\prime}_{k-1} \rangle +\langle V^{\prime}_{k-1}\rangle.
\end{multline}
Since
\begin{equation}
\langle  V^{\prime}_k\rangle=(1-q)\langle U^{\prime}_{k-1}\rangle -(1-q) \langle V^{\prime}_{k-1} \rangle,
\end{equation}
we obtain 
\begin{equation}
\langle U^{\prime}_{k-1}\rangle - \langle V^{\prime}_{k-1} \rangle = \frac{1}{1-q} \langle  V^{\prime}_k\rangle,
\end{equation}
and therefore
\begin{equation}
\langle [F^n(x)]_j \rangle =(1-q)^2 q + \langle U_{k} \rangle
- \frac{1}{1-q} \langle  V^{\prime}_k\rangle. 
\end{equation}
Now we need to find  $\langle V^{\prime}_{k} \rangle$. As noted before, the recurrence equations for 
$\langle U^{\prime}_{n} \rangle, \langle V^{\prime}_{n} \rangle$
are the same as for  $\langle U_{n} \rangle, \langle V_{n} \rangle$, that is, as in eq. (\ref{recuV}). Only initial conditions
are different,
\begin{equation}
 \langle U^{\prime}_{2} \rangle=\langle \bar{X}_1( 1-\bar{X}_1 \bar{X}_2) \rangle=
\langle \bar{X}_1 -\bar{X}_1 \bar{X}_2 \rangle
=1-q - (1-q)^2=q-q^2,
\end{equation}
\begin{equation}
 \langle V^{\prime}_{2} \rangle=\langle \bar{X}_1\bar{X}_2( 1-\bar{X}_1 \bar{X}_2 )\rangle=\langle \bar{X}_1 \bar{X}_2 -\bar{X}_1 \bar{X}_2 \rangle
=0.
\end{equation}
The formula (\ref{finalUV}) thus becomes
\begin{equation}
 \left[ \begin {array}{c} 
\langle U^{\prime}_n \rangle\\
\langle V^{\prime}_n \rangle
 \end {array} \right]
=P
\left[ \begin {array}{cc} \lambda_1^{n-2}&0
\\ \noalign{\medskip}0&\lambda_2^{n-2}\end {array} \right] P^{-1}
\left[ \begin {array}{c} 
q-q^2\\
0 
 \end {array} \right],
\end{equation}
where $P$ and $P^{-1}$ are defined in eq. (\ref{defP}).
After carrying out matrix multiplication and simplification this yields
\begin{equation}
 \langle V^{\prime}_n \rangle= \frac{q(1-q)^2}{\lambda_1-\lambda_2} \left (\lambda_1^{n-2} - \lambda_2^{n-2} \right).
\end{equation}
The final result is thus
\begin{multline}
\langle [F^n(x)]_j \rangle =(1-q)^2 q  
+ \frac{q}{\lambda_2-\lambda_1} \left( \alpha_1 \lambda_1^{k-2} + \alpha_2 \lambda_2^{k-2} \right) \\
-  \frac{q(1-q)}{\lambda_1-\lambda_2} \left (\lambda_1^{k-2} - \lambda_2^{k-2} \right),
\end{multline}
which simplifies to
\begin{multline} \label{periodic1}
\langle [F^n(x)]_j \rangle =(1-q)^2 q  \\
+ \frac{q}{\lambda_2-\lambda_1} \left( (\alpha_1+1-q) \lambda_1^{k-2} + (\alpha_2-1+q) \lambda_2^{k-2} \right),
\end{multline}
where $\lambda_{1,2}$ and $\alpha_{1,2}$ are defined in eq. (\ref{eigenvalues}).
Note that the above expression does not depend on $n$, which means that $\langle [F^n(x)]_j \rangle $ becomes constant
when $n \geq k$. 

In fact, one can show that when $n=k-1$ and $n=k-2$, eq. (\ref{periodic1}) remains valid. We will omit details, but the reasoning is very similar as
in the case of $n \geq k$. The final result for the periodic case can thus be summarized as follows.
\begin{multline} \label{finitesolution}
 \langle [F^n(x)]_j \rangle = 
(1-q)^2 q \\+
\begin{cases}
\frac{q^2}{\lambda_2-\lambda_1} \left( \alpha_1 \lambda_1^{n-1} + \alpha_2 \lambda_2^{n-1} \right) & \text{if}\,\,\, n \leq k-3, \\
\frac{q}{\lambda_2-\lambda_1} \left( (\alpha_1+1-q) \lambda_1^{k-2} + (\alpha_2-1+q) \lambda_2^{k-2} \right) & \text{if}\,\,\, n \geq k-2.
\end{cases}
\end{multline}

\section{Finite vs. infinite configurations}
The case of the periodic initial configuration is often interpreted as a finite configuration of $k$ sites with periodic boundary conditions.
We can say, therefore, that we have obtained probabilistic solutions for both infinite (eq. \ref{infinitesolution}) and finite
(eq. \ref{finitesolution}) configurations. Let us briefly describe differences between them. To simplify
notation, we will define $c_n= \langle [F^n(x)]_j \rangle$, and we will call $c_n$ the \emph{density of ones}  after $n$
iterations of the rule.
 \begin{figure}
  \begin{center}
    \includegraphics[scale=0.8]{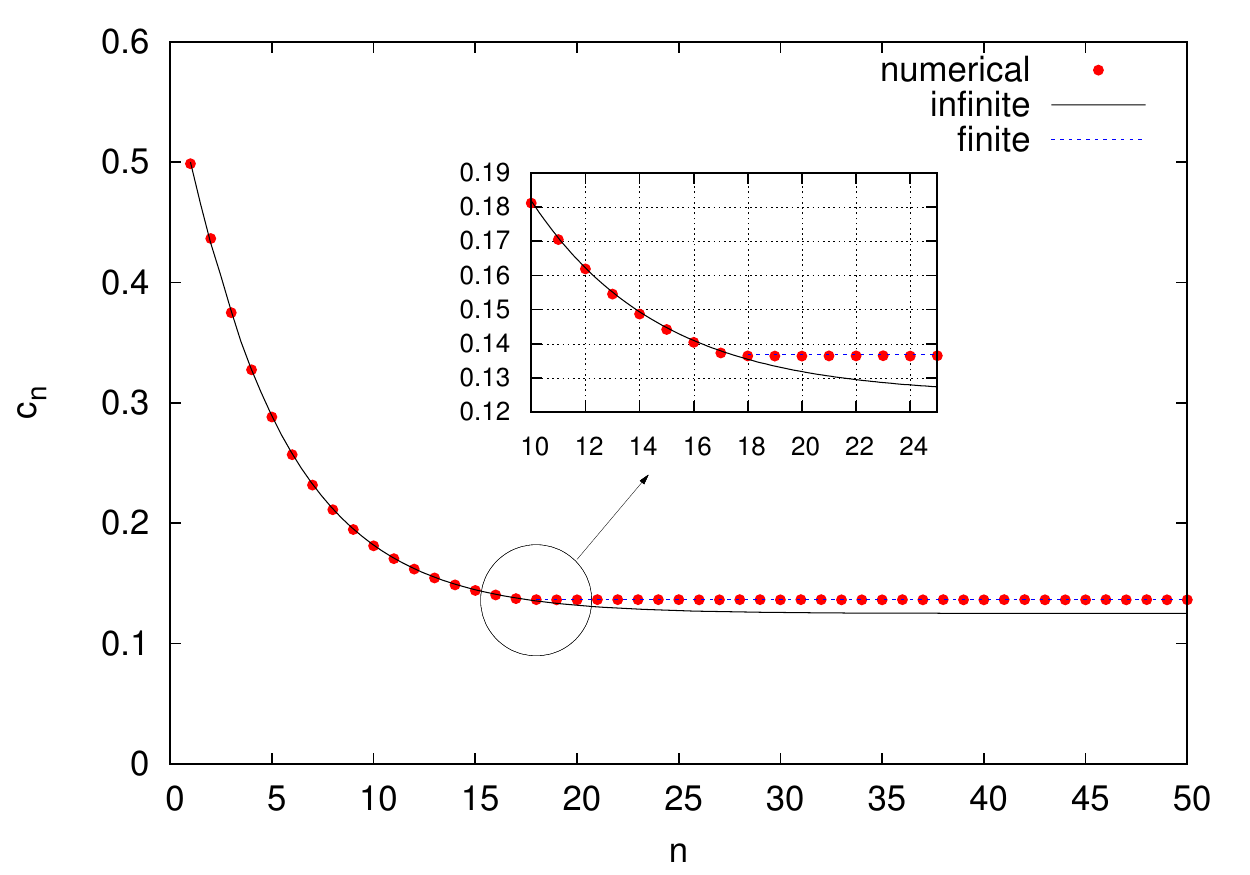}
  \end{center}
 \caption{Plot of  $c_n$ as a function of $n$ for $q=1/2$ for infinite initial configuration (solid line)
and periodic initial configuration with period $k=20$ (dashed line). Dots represent 
average value of the state of site $i=0$ after $n$ iterations of rule 172, for finite periodic configuration of $20$
sites, obtained by direct numerical iteration of randomly generated initial configuration  repeated $10^6$ times. }
  \label{densoft}
\end{figure}

\begin{figure}
  \begin{center}
    \includegraphics[scale=0.8]{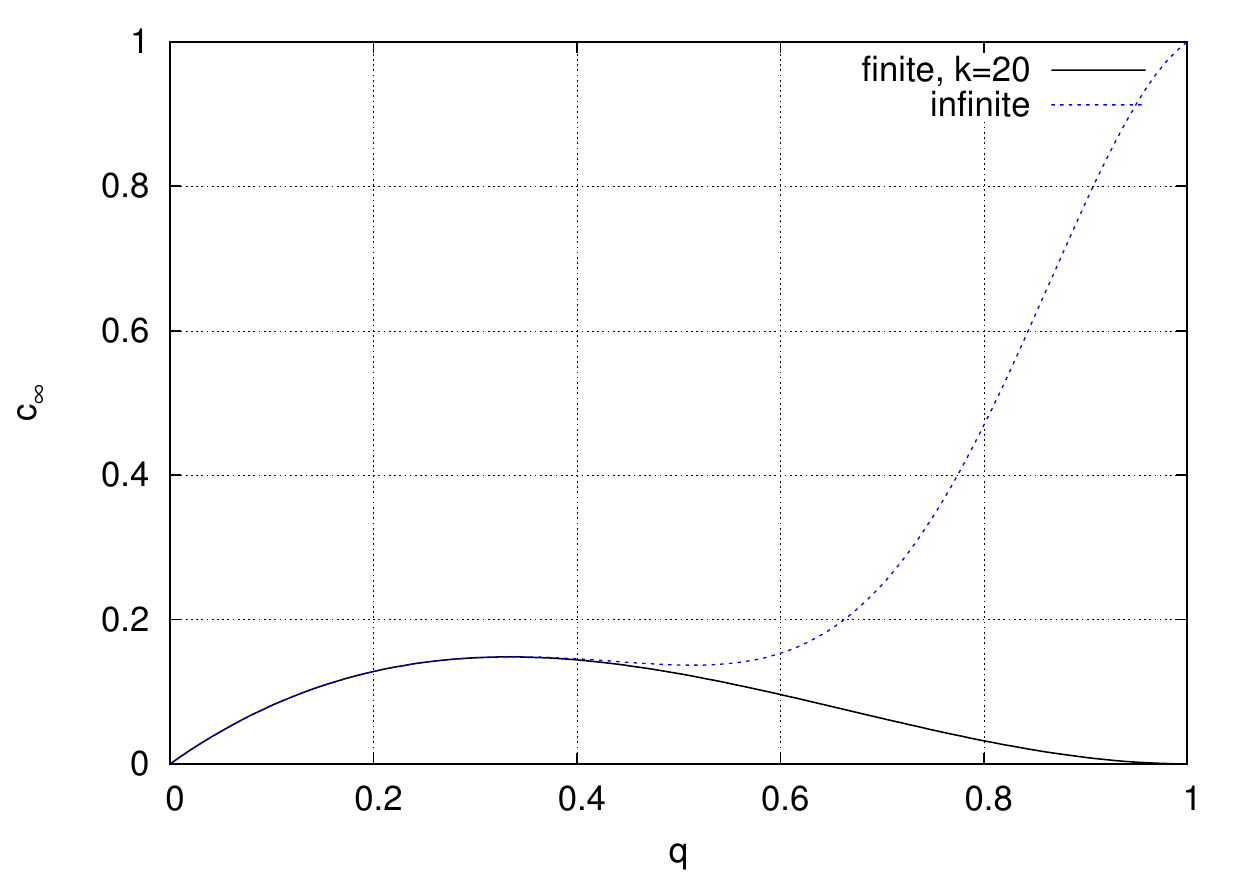}
  \end{center}
 \caption{Plot of $c_{\infty}$ as a function of~$q$ for  infinite initial configuration (solid line)
 and periodic initial configuration with period $k=20$ (dashed line).}
  \label{steadystate}
\end{figure}

Figure \ref{densoft} shows plots of the density $c_n$ versus $n$ for both finite ($k=20$) and infinite configurations
with $q=0.5$. One can see that initially they are identical, and at $n=k-2=18$ they split. The steady state density
is clearly higher in the finite case, albeit not too much. The difference between finite and infinite configurations 
is much more dramatic when the initial density $q$ becomes closer to 1. This is demonstrated Figure~\ref{steadystate}
which shows the graph of $c_{\infty}=\lim_{n \to \infty} c_n$ as a function of~$q$. One can see that the difference between 
steady states of finite and infinite configurations grows rapidly when $q$ approaches 1. In fact, in the vicinity of $q=1$,
finite configurations tend to density approaching zero, while infinite configurations tend to density approaching 1.
This shows the danger of using finite lattices with periodic boundaries as somewhat ``resembling'' infinite ones,
as it is sometimes done in cellular automata simulations and models. We can conclude from Figure~\ref{steadystate} that finite
size effects can be very significant in CA, even leading to outcomes completely opposite than those  expected
for an infinite system.

\section{Conclusions}
We have demonstrated that a simple CA rule, namely rule 172, can be explicitly solved, meaning that it is
possible to obtain a closed form formula for the state of a given cell after $n$ iterations of the rule,
as in eq. (\ref{solj}). Such formula is useful for further analysis of properties of the rule. Using
it, we obtained ``probabilistic'' solutions, that is, the expected value of a cell after $n$ iterations for
both infinite and finite configurations, assuming that the initial state is drawn from a Bernoulli
distribution. This, in turn, allowed us to investigate the role and significance of finite size effects for 
rule 172.

It should be stressed that probabilistic solution obtained here is exact, and thus it should not be confused with
approximate methods such as the mean-field theory \cite{Gutowitz95meanfield} or local structure theory
\cite{gutowitz87a,paper50}

Although the method presented here is, obviously,  not a general one, it can be used for other
rules providing that their dynamics is not overly complicated.  While it is difficult to pinpoint what
``not overly complicated'' means precisely, some empirical observations can be made. First of all, let us
notice that in rule 172 if a pair of zeros occurs somewhere in the initial condition, it stays
in the same place throughout iterations of the rule. The word 00 is thus  \emph{a blocking word} of rule
172 (see \cite{Kurka2009} for precise definition). It is known that all CA rules possessing a blocking word are
almost equicontinuous \cite{Kurka2009}, thus rule 172 has that property too. Existence of a blocking word
severely limits propagation of information between sites, thus making the rule ``simple'', and a good candidate for
solving using the method outlined in this paper. One should add, however, that the rule 172 is
only \emph{almost} equicontinuous, but not equicontinuous. Furthermore, its entropy is positive, as shown
in the appendix, thus its dynamics is not entirely trivial. The fact that it can nevertheless be solved is
encouraging, and it seems highly probable that other almost-equicontinuous rules with positive entropy
could be solved in a similar fashion. 

\vskip 1em
\noindent\textbf{Acknowledgments}\\
The author acknowledges financial support from the Natural Sciences and
Engineering Research Council of Canada (NSERC) in the form of Discovery Grant.
This work was made possible by the facilities of the Shared
Hierarchical Academic Research Computing Network (SHARCNET:
www.sharcnet.ca) and Compute/Calcul Canada.

\section*{Appendix: entropy of rule 172}
Let us recall that ${\cal A} =\{0,1\}$ and that the local function of rule 172 is defined as
 \begin{equation}
 f(x_1,x_2,x_3) = \left \{ \begin{array}{ll}
                              x_2     & \mbox{if $x_1=0$,}\\
                              x_3     & \mbox{if $x_1=1$.}
                    \end{array}
 \right.
\end{equation}
Corresponding to $f$,  we define a
 {\em global mapping}  $F:{\cal A}^\ZZ \to {\cal A}^ \ZZ$ such that
$
(F(x))_i=f(x_{i-1},x_i,x_{i+1})
$
 for any $x\in \cal A^ \ZZ$. We will be interested in the entropy of the dynamical 
system $({\cal A}^\ZZ, F)$, to be denoted by $h({\cal A}^\ZZ, F)$.

\begin{proposition}
 Entropy of $({\cal A}^\ZZ, F)$, where $F$ is the global function of CA rule 172, is positive.
\end{proposition}
Let $\Sigma_{\{00,010\}}$ be the set of all elements of ${\cal A}^\ZZ$ in which words
$00$ and $010$ do not occur. $(\Sigma_{\{00,010\}}, \sigma)$ is then a subshift of finite type,
where $\sigma$ is the usual shift map, defined as $\sigma(x)_i=x_{i+1}$. We will first show that
rule 172 restricted to $\Sigma_{\{00,010\}}$ is equivalent to shift map,
\begin{equation} \label{restriction}
F |_{\Sigma_{\{00,010\}}}=\sigma. 
 \end{equation}
Indeed, consider  the value of $f(x_{i-1},x_i,x_{i+1})$ for $x \in \Sigma_{\{00,010\}}$.
If $x_{i-1}=0$, we must have $x_i=1$ (because double zeros are forbidden), and $x_{i+1}=1$
(because isolated ones are forbidden), therefore $f(0,x_i,x_{i+1})=f(0,1,1)=1=x_{i+1}$.
If, on the other hand,  $x_{i-1}=1$, then by the  definition of $f$ for rule 172 shown
 in eq. (\ref{selector}), 
$f(1,x_i,x_{i+1})=x_{i+1}$. This means that $f(x_{i-1},x_i,x_{i+1})=x_{i+1}$ for all $x \in \Sigma_{\{00,010\}}$,
as required.

Let us now compute the entropy of $(\Sigma_{\{00,010\}}, \sigma)$ using the method
outlined in \cite{Lind95}. First, 
we need to find a Markov shift
conjugate to $(\Sigma_{\{00,010\}}, \sigma)$. This can be done by defining new symbol set 
${\cal B}=\{011,101,111 \}:=\{a,b,c\}$. If $\Sigma_{\{ba\}}$ denotes the set of points of ${\cal B}^\ZZ$ in
which the word $ba$ does not occur, then it is easy to show that the subshift
 $(\Sigma_{\{00,010\}}, \sigma)$ is
conjugate to  $(\Sigma_{\{ba\}}, \sigma)$, which is a Markov subshift.
Adjacency matrix $3 \times 3$ for $(\Sigma_{\{ba\}}, \sigma)$ is defined by $M_{i,j}=1$ if $(i,j) \neq (b,a)$ 
and $M_{i,j}=0$ if $(i,j) = (b,a)$. Spectral radius of this matrix is $(3+\sqrt{5})/{2}$, therefore
\begin{equation}
 h\left(\Sigma_{\{ba\}}, \sigma\right)=h\left(\Sigma_{\{00,010\}}, \sigma\right)=
h\left(\Sigma_{\{00,010\}}, F\right)=\ln \frac{3+\sqrt{5}}{2},
\end{equation}
where we used the fact that the entropy is invariant with respect to conjugacy.
This shows that $h\left(\Sigma_{\{00,010\}}, F\right)>0$. Since $\Sigma_{\{00,010\}} \subset {\cal A}^\ZZ$,
we conclude that $h\left({\cal A}^\ZZ, F\right)>0$.
\bibliographystyle{hfbib}
\bibliography{extracted}
\end{document}